\documentclass[pmlr]{jmlr}

\usepackage{longtable}

\usepackage{booktabs}
\usepackage[load-configurations=version-1]{siunitx} 

\usepackage{hyperref}
\usepackage{amsmath,amssymb}
\usepackage{mathrsfs}
\usepackage{natbib}

\usepackage{float}

\newsavebox\CBox
\def\textBF#1{\sbox\CBox{#1}\resizebox{\wd\CBox}{\ht\CBox}{\textbf{#1}}}

\usepackage{multirow}
\usepackage{booktabs}
\usepackage[outline]{contour}
\usepackage{xcolor}
\contourlength{0.15pt}
\contournumber{10}

\theorembodyfont{\upshape}
\theoremheaderfont{\scshape}
\theorempostheader{:}
\theoremsep{\newline}

\jmlrvolume{166}
\jmlryear{2022}
\firstpageno{90}
\jmlrworkshop{HEAR: Holistic Evaluation of Audio Representations}

\title[The Efficacy of Self-Supervised Speech Models as Audio Representations]{The Efficacy of Self-Supervised Speech Models as Audio Representations}

\usepackage[symbol]{footmisc}

\author{
\Name{Tung-Yu Wu}\thanks{with equal contribution} \Email{b08901133@ntu.edu.tw} \\
\addr Department of Electrical Engineering \\
National Taiwan University
\AND
\Name{Tsu-Yuan Hsu}\footnotemark[1] \Email{b08201047@ntu.edu.tw} \\
\addr Department of Computer Science and Information Engineering \\
National Taiwan University
\AND
\Name{Chen-An Li}\footnotemark[1] \Email{b08902123@ntu.edu.tw} \\
\addr Department of Computer Science and Information Engineering \\
National Taiwan University
\AND
\Name{Tzu-Han Lin}\footnotemark[1] \Email{b08902126@ntu.edu.tw} \\
\addr Department of Computer Science and Information Engineering \\
National Taiwan University
\AND
\Name{Hung-yi Lee} \Email{hungyilee@ntu.edu.tw} \\
\addr Department of Electrical Engineering \\
National Taiwan University
}

\editors{Joseph Turian, Björn W. Schuller, Dorien Herremans, Katrin Kirchhoff, Paola Garcia Perera, Philippe Esling}

\begin{document}
\graphicspath{ {./images/} }
\maketitle

\begin{abstract}
Self-supervised learning (SSL) speech models, which can serve as powerful upstream models to extract meaningful speech representations, have achieved unprecedented success in speech representation learning. However, their effectiveness on non-speech datasets is relatively less explored. In this work, we propose an ensemble framework, with a combination of ensemble techniques, to fuse SSL speech models’ embeddings. Extensive experiments on speech and non-speech audio datasets are conducted to investigate the representation abilities of our ensemble method and its single constituent model. Ablation studies are carried out to evaluate the performances of different ensemble techniques, such as feature averaging and concatenation. All experiments are conducted during NeurIPS 2021 HEAR Challenge as a standard evaluation pipeline provided by competition officials. Results demonstrate SSL speech models' strong abilities on various non-speech tasks, while we also note that they fail to deal with fine-grained music tasks, such as pitch classification and note onset detection. In addition, feature ensemble is shown to have great potential on producing more holistic representations, as our proposed framework generally surpasses state-of-the-art SSL speech/audio models and has superior performance on various datasets compared with other teams in HEAR Challenge. Our code is available at https://github.com/tony10101105/HEAR-2021- NeurIPS-Challenge—NTU-GURA.
\end{abstract}

\section{Introduction}
\label{sec:intro}
Data representation is crucial to the performance of deep learning algorithms. A good representation catches the underlying patterns of the input data \citep{representationLearning} and can serve as effective inputs for neural networks, which usually reach better performance with a higher quality of inputs. The extraction of data can be done either in a hand-crafted way using heuristic methods or with the help of neural networks. For neural-network-based feature extractions, supervised learning, though with a clearer training objective, has limited scalability, while unsupervised learning needs only unlabeled data but usually with inferior results.

In recent years, self-supervised learning (SSL) frameworks for speech representation, such as HuBERT \citep{hubert} and wav2vec 2.0 \citep{wav2vec2}, have proved their powerful abilities to extract useful speech audio features and deal with a wide range of downstream tasks, such as Speaker Identification (SID) and Automatic Speaker Verification (ASV). The achievements are mainly acquired by the use of various pre-training techniques with self-defined tasks on a huge amount of unlabeled corpus. For example, during the pre-training stage, HuBERT is required to predict the frame-level cluster assignments of the masked part of the input sequence. By solving self-defined tasks, SSL speech models are capable of capturing meaningful latent representations of given speech audio clips to help with diverse downstream tasks.

Former researches about SSL speech models mainly focus on their representation abilities on speech corpus, while their effectiveness on non-speech tasks still lacks exploration. Also, given that they are all powerful feature extractors and features generated by each of them may be distinct due to different pre-training techniques, studies about speech representation ensemble are not abundant. Therefore, this work proposes a speech representation ensemble framework that integrates several models' output features with aggregation and concatenation techniques. Our framework is evaluated on the official benchmark of NeurIPS 2021 HEAR Challenge \citep{pmlr-v176-turian22a}, which contains 16 speech and non-speech datasets, to investigate both SSL models' representation abilities on audio data and the feasibility of speech representation ensemble. We found that wav2vec 2.0 and HuBERT are capable of tackling some tasks that are unrelated to their pretraining corpus. For example, HuBERT acquires the second-best result among all submitted single models on Gunshot \citep{gunshots} dataset, which is a sound distance classification task. This indicates HuBERT's capacity of preserving the information of distance from the sound source. However, we also note that, though wav2vec 2.0 and HuBERT do well on instrument sound classification, their performances on fine-grained music tasks such as pitch classification drop drastically. We also observe the performance gaps between wav2vec 2.0 and HuBERT on various speech and non-speech datasets (the former may have high-quality feature presentations on certain datasets, where the latter cannot extract well, and vice versa). Finally, we demonstrate that, with appropriate representation ensemble strategies, the fusion of different model embeddings, with more holistic information, can achieve higher performances, and our ensemble framework indeed displays competitive results among wav2vec 2.0, HuBERT, and other teams' approaches.

In summary, the contributions of this work are twofold:
\begin{enumerate}
    \item Investigating SSL speech models' effectiveness on diverse audio datasets. We found that current SOTA SSL speech models (wav2vec 2.0 and HuBERT) are capable of some non-speech scenarios, such as instrument sound classification and environmental sound event detection. However, failure on the pitch- and note-related tasks is also observed. In addition, we found the performance gaps between wav2vec 2.0 and HuBERT on several datasets.

    \item Proposing a speech representation ensemble framework for integrating SSL speech representation models. Experimental results on HEAR benchmark demonstrate its superiority over state-of-the-art methods and ensemble learning's potential in the field of speech representation learning.
\end{enumerate}

\section{Related Works}\label{sec:relatedWork}

\subsection{SSL speech and audio Models}
The objective of SSL is to extract meaningful representations from the input data without any human annotations. It arranges auxiliary tasks with unlabeled data, driving models to learn the representative features by solving these tasks. Models pre-trained through SSL skills on diverse large corpus can be used as effective feature extractors afterward. As a whole, pre-training strategies for speech/audio models can be roughly categorized into three groups: generative, discriminative, and multi-task learning.

Generative training investigates the distribution of input data. Given a sequence of samples, models are asked to generate new samples to fit into the original distribution. APC \citep{APC} is pre-trained with a unidirectional RNN in an autoregressive manner to predict the information of future frames. VQ-APC \citep{VQ-APC} extends APC by applying Vector Quantized (VQ) layers to generate better quantized representation. PANNs \citep{PANNs}, which is an SSL audio model pre-trained on AudioSet \citep{AudioSet}, proposes Wavegram-Logmel-CNN to concatenate both waveform and the log-mel spectrogram as input representations. PaSST \citep{PaSST} presents the patchout method to optimize Transformers with audio spectrograms. Wav2CLIP \citep{Wav2clip} distills representations from CLIP \citep{CLIP} to produce general and robust audio representations.


Discriminative training aims to discriminate positive samples from negative samples in the embedding space. CPC \citep{CPC} extracts features that maximally preserve the mutual information of the input sequence and the context latent representations over the time horizon. It makes use of a non-linear CNN-based encoder to embed the input sequence into latent representations and an autoregressive RNN-based network to generate the context latent representations by summarizing the encoder's previous timesteps. Wav2vec \citep{wav2vec} improves CPC by adopting a CNN-based autoregressive network to parallelize the training process and thus enhance time efficiency. vq-wav2vec \citep{vq-wav2vec} further boosts wav2vec by introducing VQ modules, such as Gumbel-Softmax \citep{gumbel-softmax} and online k-means clustering, to learn discrete speech representations. wav2vec 2.0 refines the framework of vq-wav2vec by merging the two-stage pipeline to perform end-to-end joint training. HuBERT is pre-trained on the task of predicting the k-means cluster of the masked tokens, currently achieving state-of-the-art performance on many speech tasks.


Multi-task learning solves multiple speech/audio tasks simultaneously to guide models to generate more comprehensive and robust representations. PASE \citep{PASE} utilizes regressors to predict training objectives, such as waveform, log power spectrum (LPS), mel-frequency cepstral coefficients (MFCC), and prosody features. Discriminators are also introduced into PASE, and strategies to sample positive and negative samples, such as local info max (LIM), global info max (GIM), and sequence predicting coding (SPC) are adopted. PASE+ \citep{PASE+} improves PASE by applying online speech distortion modules to add several noises to input speech and using Quasi-RNN (QRNN) \citep{QRNN} to help the encoder learn the long-term dependencies. 


\subsection{Ensemble Learning}
Ensemble learning is a machine learning technique that aggregates diverse base models to form an ensemble system with better robustness and performance. Generally, ensemble strategies can be categorized into boosting, bagging, and stacking \citep{ensemble}. Boosting strategy integrates weak models, converting them into a new learner with better generalization. Bagging strategy works by training several weak models on datasets where each data point is randomly sampled from the original dataset, aiming to reduce model variance. Stacking strategy takes predictions generated by each model as input, learning how to combine input predictions to achieve better performance. For the feature ensemble of SSL speech representation models, there are some intuitive operations \citep{arunkumar2022investigation} such as feature summation, averaging, concatenation, and fusing with attention layers. It has been shown that feature ensemble achieves higher performance in automatic speaker recognition (ASR) \citep{arunkumar2022investigation}.

\section{Methods}\label{sec:method}
We argue that speech models with different SSL techniques produce inhomogeneous feature representations given the same audio input. Hence, even though these cutting-edge models' performances on various benchmark datasets are close, it could still be beneficial to merge their representations for climbing up to a higher performance. Hence, we propose an ensemble framework that integrates the representations of speech models, which is further shown to be more holistic and contain more information compared with a single SSL speech/audio model.

As shown in \figureref{fig:modelArchitecture}, the advocated ensemble framework comprises three models and two ensemble techniques. The three models are wav2vec 2.0, HuBERT, and CREPE \citep{CREPE}. The two techniques are feature aggregation and feature concatenation, where the former is an intra-model operation that averages the output of a network's different layers to form a single output feature and the latter is an inter-model operation that combines models' output features as the final representation of input audio.

\begin{figure}[htbp]
\floatconts
  {fig:modelArchitecture}
  {
  \caption{The model architecture of the proposed framework. We first take the average of each layer's feature. Next, linear interpolation is applied to align each constituent model's averaged output. Finally, representations are concatenated to form the final representation.}
  }
  {\includegraphics[width=0.9\linewidth]{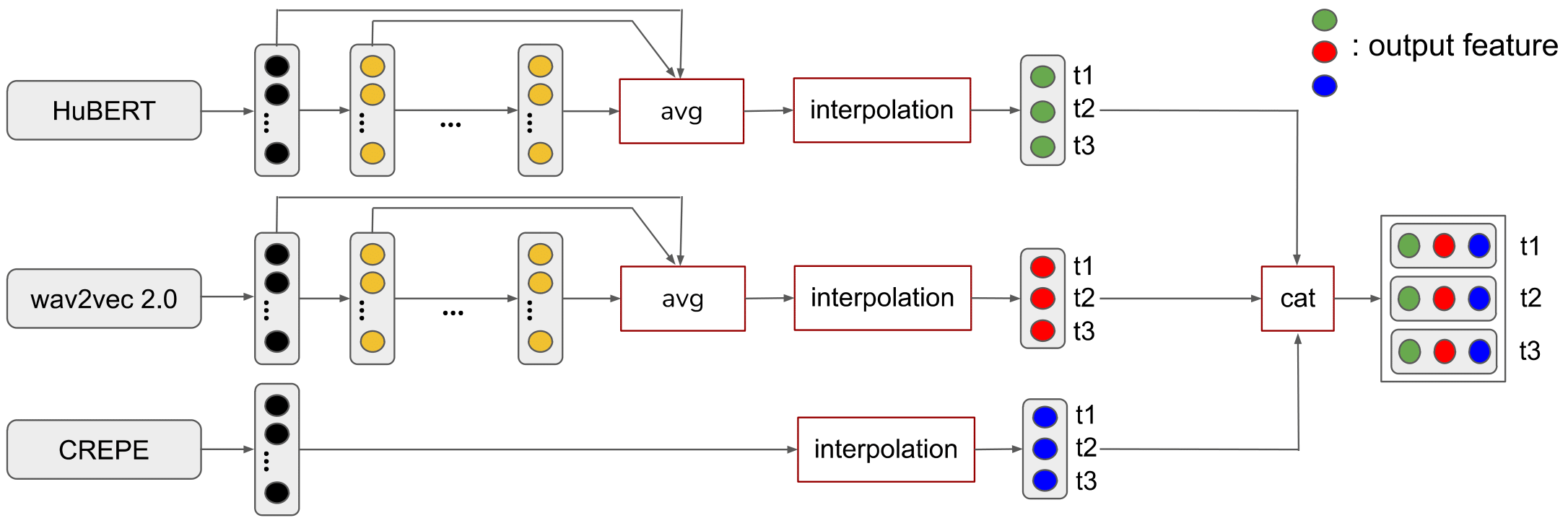}}
\end{figure}

\subsection{Models for Ensemble}
Although the number of models for the bagging method is unlimited, three or five are often chosen in practice for balancing size and performance. Our method in HEAR adopts three models: wav2vec 2.0, HuBERT, and CREPE, which are trained on different datasets, paradigms, and techniques. Wav2vec 2.0 and HuBERT are both SSL speech models, while the former is pre-trained with masked vector quantization plus contrastive discrimination, and the latter is masked vector quantization plus token prediction/classification. Wav2vec 2.0 has Base and Large versions, and HuBERT has Base, Large, and Extra Large versions. The Base versions of both models are pre-trained on LibriSpeech 960 hr \citep{LibriSpeech}, while other versions are trained on Libri-Light 60k hr \citep{Libri-Light}. CREPE is a CNN-based pitch estimation model pre-trained on several pitch-based datasets, such as RWC-Synth \citep{rwc-synth}, MDB-STEM-Synth \citep{MDB-STEM-Synth} and NSynth, with supervised training manner. Following its official implementation, we use CREPE's fifth max-pooling layer output as its feature representation of input audio. The adoption of CREPE is based on the observation that HuBERT and wav2vec 2.0 seem unable to handle pitch-related tasks well, which is further discussed in the experiment \sectionref{sec:results}.

It should be emphasized that our framework does not limit the number and types of models to be assembled. It is completely plausible to replace wav2vec 2.0, HuBERT, and CREPE with other models, such as vq-wav2vec and TERA, while only the three models are studied during the challenge.


\subsection{Feature Aggregation}
Each layer of SSL Transformer-based speech models may have attention to a certain aspect of sequential input data. As a result, instead of merely using the last hidden state of the model, we take the average of all the hidden states, i.e., fuse the output of each transformer block plus the initial embedding outputs, to yield more comprehensive feature representations, as shown in \figureref{fig:modelArchitecture}. There have been various layer aggregation methods \citep{aggregation} studied, while we simply utilize the most straightforward one for simplicity. In other words, though the choice and weight of hidden states are both optional, we sum up all the hidden states fairly. Our aggregation operation can be generally defined as:
\begin{equation} \label{eq:1}
    X = \frac{1}{L}\sum_{i=1}^{L} X_i,
\end{equation}
where $L$ is the number of Transformer blocks and $X_i$ is the $i$-th Transformer block's output. Notice that the dimensions of different hidden states are all identical in both wav2vec 2.0 and HuBERT, so we can directly take the average, as shown in \equationref{eq:1}.

It should be noted that feature aggregation is not applied to CREPE since it is a relatively shallow vanilla CNN model with six convolutional layers. We only leverage the output of the fifth max-pooling layer as the feature embedding, following its official implementation\footnote{https://github.com/marl/crepe}.

\subsection{Feature Concatenation}
The dimension of a representation model's output features can be roughly described as $T*C$, where $T$ is the timestamp dimension and $C$ is the feature dimension of each timestamp. Since HuBERT, wav2vec 2.0, and CREPE have different output dimensions (the output dimension of HuBERT and wav2vec 2.0 depends on their versions, such as wav2vec 2.0 Base or wav2vec 2.0 Large), up/downsampling is required for feature concatenation. For simplicity, we linearly interpolate the timestamp dimension $T$ and feature dimension $C$ of each constituent model' representation into those of wav2vec 2.0. Next, features are concatenated along the feature dimension so that the dimension of the final output feature is $T_w*(3C_w)$, where the subscript $w$ denotes wav2vec 2.0. This process preserves the time information of the final output feature, which is vital to certain downstream tasks such as sound event detection.

\section{Experimental Setup}\label{sec:expSetup}
Experiments are carried out during the NeurIPS 2021 HEAR Challenge, where 16 iconic or novel datasets are used for the holistic evaluation of audio representations. HEAR officials conduct the evaluation process, and the full results are displayed on its official benchmark website.\footnote{https://hearbenchmark.com}

\begin{table}[hbtp]
    \centering
    \caption{Categories of 16 datasets presented in HEAR Challenge's evaluation process.}\label{Tab:1}
    \tiny
    \resizebox{\columnwidth}{!}{
    \begin{tabular}{c c}
        \toprule
        Category & Datasets\\
        
        \midrule
        \multirow{5}{*}{Speech}
        & CREMA-D \citep{CREMA-D}\\
        & LibriCount \citep{libricount}\\
        & Speech Commands \citep{SpeechCommand}\\
        & VoxLingua107 \citep{voxlingua107}\\
        & Vocal Imitation Set \citep{vocalImitionSet}\\
        
        \midrule
        \multirow{4}{*}{Instrument}
        & Nsynth \citep{NSynth}\\
        & MAESTRO \citep{MAESTRO}\\
        & Beijing Opera Percussion \citep{BeijingOpera}\\
        & Mridingham Stroke and Tonic \citep{MridangamStroke}\\
        
        \midrule
        \multirow{4}{*}{Environmental Sounds}
        & ESC-50 \citep{esc-50}\\
        & Beehive States \citep{beehive}\\
        & DCASE 2016 Task 2 \citep{DCASE}\\
        & Gunshot Triangulation \citep{gunshots}\\
        
        \midrule
        \multirow{3}{*}{General}
        & FSD50k \citep{fsd50k}\\
        & GTZAN Music/Speech \citep{GTZANMusicSpeech}\\
        & GTZAN Genre Collection \citep{GTZAN-genre}\\
        
        \bottomrule
    \end{tabular}
    }
\end{table}

\subsection{HEAR Challenge and Evaluation Pipeline}
HEAR Challenge is one of the challenges in NeurIPS 2021 which aims at developing audio representation frameworks with high generalizability. HEAR has established a comprehensive benchmark that currently includes 19 downstream tasks, including but not limited to speech, environmental sound, and music domains. Our experiments follow HEAR's evaluation pipeline, which adopts the evaluation principles of representation quality proposed by \citep{eval}. In brief, HEAR evaluates the effectiveness of submitted upstream models by investigating the performances of downstream shallow models, trained with embeddings generated by upstream models, on each task. Since HEAR's philosophy is to foster the development of audio representation methods with strong generalizability, models are anticipated to have decent performances on all datasets, rather than excellent results on several datasets but poor on other tasks.

Notably, HEAR requires challenges to provide two representations: timestamp embedding and scene embedding. The former is the representation that has one dimension as the time dimension, and the latter is a single embedding that represents the whole audio. A baseline method to generate scene embedding is to take the average of timestamp embedding along the time dimension. This work adopts this approach during the challenge.

\subsection{Datasets}
HEAR benchmark currently contains 19 tasks with 16 datasets. Some tasks utilize the same datasets. Datasets adopted in HEAR include Speech Commands\citep{SpeechCommand}, NSynth \citep{NSynth}, DCASE 2016 Task 2 \citep{DCASE}, Beehive States \citep{beehive}, Beijing Opera Percussion Instrument \citep{BeijingOpera}, CREMA-D \citep{CREMA-D}, ESC-50 \citep{esc-50}, FSD50k \citep{fsd50k}, Gunshots recorded in an open field using iPod Touch devices (Gunshot Triangulation) \citep{gunshots}, GTZAN Genre Collection \citep{GTZAN-genre}, GTZAN Music/Speech \citep{GTZANMusicSpeech}, LibriCount \citep{libricount}, MAESTRO \citep{MAESTRO}, Mridangam Stroke \citep{MridangamStroke}, Vocal Imitation Set \citep{vocalImitionSet} and VoxLingua107 \citep{voxlingua107}.

Categories of these 16 datasets are presented in \tableref{Tab:1}. Speech Command is a speech command classification dataset that contains 105829 utterances of 35 words. NSynth is a pitch classification dataset from 1006 instruments. DCASE 2016 Task 2 is an office sound event detection dataset with 11 event categories, such as clearing throat and coughing. It should be noticed that HEAR uses a different way to split this dataset, so the evaluation result is not comparable to other work. Beehive States is a binary beehive sound classification dataset with 930 sound clips. Beijing Opera Percussion Instrument is a percussion instrument classification dataset with 6 percussion instruments, such as Ban and Gong, which are organized into four classes, such as Bangu and Daluo. CREMA-D is a 6-class audiovisual emotion recognition dataset. HEAR only adopts its audio recordings. ESC-50 is a 50-class environmental sounds classification dataset. FSD50K is a 200-class sound event classification dataset with 51197 Freesound \citep{freesound} clips drawn from AudioSet. Gunshot Triangulation is a sound source distance classification dataset. GTZAN Genre Collection is a 10-class music genre classification dataset. GTZAN Music/Speech is a speech-or-music binary classification dataset. LibriCount is a speaker number estimation dataset. MAESTRO is a note onset detection dataset. The Mridangam Stroke dataset contains 6977 audio examples of 10 different strokes played on Mridangams with 6 tonics. HEAR evaluation divides the dataset into two sub-tasks: stroke classification and tonic classification. The Vocal Imitation Set is a sound classification dataset. Models should predict which reference sound the audio is imitating. VoxLingua107 is a spoken-language classification dataset. HEAR adopts the 10-language subset.

\subsection{Our Method and Baselines}
Our ensemble framework adopts pre-trained wav2vec 2.0, HuBERT, and CREPE without any fine-tuning. As a result, the utilized wav2vec 2.0 and HuBERT are both pure speech models pre-trained on LibriSpeech or Libri-Light, while CREPE is a pure music-based model pre-trained on several pitch-related datasets.

Due to feature concatenation, our proposed ensemble method generates longer scene embeddings than the single SSL model. This affects the architecture of interfaced downstream model since its input layer should match the size of input scene embeddings. Hence, it is necessary to ensure our framework does not gain benefits from the architectures of downstream models. To investigate this issue, we design a variant of the baseline scene embeddings generation approach that first splits the timestamp embeddings along the time dimension into $k$ groups, takes the mean along the time dimension, and concatenates each sub-scene embeddings as final scene embeddings. Thus, the length of scene embeddings will be $(k*C_w)$ rather than $C_w$. With this pipeline, the scene embeddings are much longer, while the representation information of the three models will still be preserved. The group number $k$ is set to 5.

Baselines include vanilla wav2vec 2.0, HuBERT, CREPE, and the other five teams' methods. For vanilla wav2vec 2.0 and HuBERT, only the last hidden state is used as the input embedding. The five teams' models are officially considered to have the most competitive performances out of all participants. Five models include Wav2CLIP \citep{Wav2clip}, PaSST \citep{PaSST}, PANNs \citep{PANNs}, EfficientNet-B2 \citep{efficientnet} and BYOL-S \citep{serab}. The first three are SSL speech and audio models, while EfficientNet is a series of CNN-based models whose architectures are determined by neural architecture search (NAS), denoted as EfficientNet-B0 to EfficientNet-B7. BYOL-S is proposed in SERAB \citep{serab} benchmark as a speech version of BYOL \citep{BYOL}. BYOL is an image representation learning framework that comprises two networks with the same architectures, referred to as the online network and target network. By taking an image as the online network's input and the augmented view as the target network's input, BYOL aims to minimize the distance between the output embeddings of these two networks. For the type of training corpus, Wav2CLIP is pre-trained on VGG-Sound \citep{vggsound}. PaSST, BYOL-S, and PANNs are pre-trained on AudioSet. EfficientNet-B2 is pre-trained on ImageNet \citep{imagenet}. Notably, each team' model implementation may be slightly different from the original paper of their methods.

\section{Results}\label{sec:results}
\subsection{Quantitative Results}

The results of 18 tasks containing 15 out of 16 datasets are displayed in \tableref{Tab:2}. We exclude the Beehive States dataset since it has a large input size which causes most teams' models to run out of memory under the competition hardware setting and get no result. The highest and second-highest scores in each dataset are marked in bold and underlined respectively.

\begin{table}[hbtp]

\caption{Performance of our framework and other baselines on 18 tasks. Some datasets are only partially evaluated, as described in \sectionref{sec:expSetup}. Our method is denoted as fusion\_cat\_xwc, which means HuBERT xLarge plus wav2vec 2.0 Large plus CREPE with intra-model feature aggregation and inter-model feature concatenation. The group-based scene embedding generation technique with our framework is fusion\_cat\_xwc\_g. Note that fusion\_cat\_xwc generally outperforms all of its constituent models and has competitive performances among all teams' approaches.} \label{Tab:2}
\resizebox{\textwidth}{!}{

  \begin{tabular}{  l | c | c | c | c | c | c | c | c | c  }

    \toprule

    \multicolumn{1}{r|}{\rotatebox{90}{\textbf{Task and measure}}} & \rotatebox{90}{\shortstack[l]{Bejing Opera\\ (Acc. $\uparrow$)}} & \rotatebox{90}{\shortstack[l]{CREMA-D\\(Acc. $\uparrow$)}} & \rotatebox{90}{\shortstack[l]{DCASE 2016\\(Onset FMS $\uparrow$)}} & \rotatebox{90}{\shortstack[l]{ESC-50\\(Acc. $\uparrow$)}} & \rotatebox{90}{\shortstack[l]{FSD50k\\(mAP $\uparrow$)}} & \rotatebox{90}{\shortstack[l]{GTZAN Genre\\(Acc. $\uparrow$)}} & \rotatebox{90}{\shortstack[l]{GTZAN Music/Speech\\(Acc. $\uparrow$)}} & \rotatebox{90}{\shortstack[l]{Gunshot\\(Acc. $\uparrow$)}} & \rotatebox{90}{\shortstack[l]{Libricount\\(Acc. $\uparrow$)}} \\



    \midrule
    
    PaSST base2levelmel \citep{PaSST} & \textBF{96.6} & {61.0} & \textBF{92.5} & \textBF{94.7} & \textBF{64.1} & \textBF{88.3} & \underline{97.7} & \textBF{94.0} & {66.0} \\
    PANNS \citep{PANNs} & {91.1} & {55.5} & {00.0} & {90.9} & {-} & {86.0} & \textBF{99.2} & {79.8} & {65.2} \\
    Wav2CLIP \citep{Wav2clip} & {93.6} & {51.2} & {00.0} & {75.9} & {36.2} & {74.8} & {94.6} & {92.9} & {52.8} \\
    SERAB BYOL-S \citep{serab} & {95.3} & {65.7} & {64.2} & {80.5} & {50.9} & {83.7} & {93.8} & {85.7} & \textBF{78.5} \\
    EfficientNet-B2 \citep{efficientnet} & {95.3} & {57.5} & {79.0} & \underline{93.5} & \underline{60.7} & \underline{87.8} & {96.8} & {87.8} & {65.1} \\
    CREPE \citep{CREPE} & {92.8} & {38.3} & {50.4} & {30.0} & {15.9} & {64.5} & {92.9} & {86.3} & {49.9} \\
    wav2vec 2.0 \citep{wav2vec2} & {90.7} & {65.6} & {66.3} & {56.1} & {34.2} & {78.0} & {94.6} & {84.8} & {69.2} \\
    HuBERT xLarge \citep{hubert} & {94.5} & {69.0} & {58.4} & {60.3} & {31.4} & {73.5} & {91.3} & {93.2} & {64.6} \\

    \midrule
    
    fusion\_cat\_xwc\_g & \underline{96.2} & \underline{74.3} & \underline{82.6} & {65.3} & {37.4} & {76.0} & {94.4} & {90.5} & {65.9} \\
    fusion\_cat\_xwc & \textBF{96.6} & \textBF{74.7} & \underline{82.6} & {73.4} & {42.0} & {80.5} & {92.8} & \underline{93.5} & \underline{69.7} \\
        
    \bottomrule

    \toprule

    \multicolumn{1}{r|}{\rotatebox{90}{\textbf{Task and measure}}} & \rotatebox{90}{\shortstack[l]{Maestro 5h\\
    (Onset FMS $\uparrow$)}} & \rotatebox{90}{\shortstack[l]{Mridingham Stroke\\
    (Acc. $\uparrow$)}} & \rotatebox{90}{\shortstack[l]{Mridingham Tonic\\
    (Acc. $\uparrow$)}} & \rotatebox{90}{\shortstack[l]{NSynth Pitch 50h\\
    (Pitch Acc. $\uparrow$)}} & \rotatebox{90}{\shortstack[l]{NSynth Pitch 5h\\
    (Pitch Acc. $\uparrow$)}} & \rotatebox{90}{\shortstack[l]{Speech commands 5h\\
    (Acc. $\uparrow$)}} & \rotatebox{90}{\shortstack[l]{Speech commands full\\
    (Acc. $\uparrow$)}} & \rotatebox{90}{\shortstack[l]{Vocal Imitation\\
    (mAP $\uparrow$)}} & \rotatebox{90}{\shortstack[l]{VoxLingua107 top 10\\
    (Acc. $\uparrow$)}} \\



    \midrule

    PaSST base2levelmel \citep{PaSST} & {-} & {96.5} & {81.9} & {54.1} & {25.6} & {68.1} & {63.9} & {18.2} & {25.9} \\
    PANNS \citep{PANNs} & {00.0} & {93.9} & {82.4} & {30.1} & {14.8} & {56.0} & {61.8} & {12.7} & {24.4}  \\
    Wav2CLIP \citep{Wav2clip} & {00.0} & {94.7} & {82.9} & {43.9} & {23.0} & {31.6} & {34.7} & {08.3} & {19.2}  \\
    SERAB BYOL-S \citep{serab} & {00.8} & \underline{97.3} & \textBF{92.8} & {71.2} & {39.6} & {91.4} & {94.8} & {16.0} & {45.8}  \\
    EfficientNet-B2 \citep{efficientnet} & {00.0} & {94.9} & {84.3} & {39.1} & {16.8} & {57.3} & {67.6} & {13.8} & {25.5}  \\
    CREPE \citep{CREPE} & \textBF{40.1} & {89.8} & {82.4} & \textBF{90.0} & \textBF{87.0} & {18.0} & {21.1} & {05.1} & {14.2} \\
    wav2vec 2.0 \citep{wav2vec2} & {03.3} & {94.3} & {82.8} & {65.3} & {40.2} & {83.8} & {87.9} & {08.0} & {49.3}  \\
    HuBERT xLarge \citep{hubert} & {00.7} & {95.3} & {85.0} & {42.9} & {18.4} & \underline{95.3} & \underline{95.4} & {15.4} & \underline{63.7}  \\

    \midrule
    
    fusion\_cat\_xwc\_g & \textBF{44.1} & \textBF{97.5} & \underline{92.4} & \underline{89.1} & \underline{85.4} & {95.1} & \textBF{96.8} & \textBF{21.5} & {62.9}  \\
    fusion\_cat\_xwc & \textBF{44.1} & {97.2} & {92.3} & {88.5} & {84.6} & \textBF{96.1} & \textBF{96.8} & \underline{19.7} & \textBF{72.0} \\
    
    \bottomrule
      
  \end{tabular}
}
\end{table}

It is demonstrated that fusion\_cat\_xwc possesses competitive performances on a large portion of datasets presented in \tableref{Tab:1}. Specifically, our method achieves top performances on Beijing Opera Percussion, CREMA-D, MAESTRO 5h, Mridingham Stroke, Speech Commands 5h, Speech Commands full, Vocal Imitations, and VoxLingua107, along with second places on DCASE 2016 Task 2, Gunshot Triangulation, LibriCount, Mridingham Tonic, NSynth 5h, and NSynth 50h. The datasets our method does not reach the top two results are ESC-50, FSD50k, GTZAN Genre, and GTZAN Music/Speech. It can also be observed that our ensemble method outperforms the state-of-the-art HuBERT xLarge and wav2vec 2.0 Large alone on all datasets except GTZAN Music/Speech and LibriCount, indicating that the ensemble framework can indeed stably increase a single model's speech/audio representation ability by incorporating multiple representation models. Furthermore, the fusion\_cat\_xwc\_g with much longer scene embeddings does not surpass the performance of the original fusion\_cat\_xwc. Specifically, it gains lower performances on 9 tasks, slightly higher results on 6 tasks, and 3 ties on tasks that only require timestamp embeddings. This shows that longer scene embeddings do not guarantee a stronger downstream model and thus higher downstream performances.

In terms of other submitted methods, PaSST base2levelmel owns first place on 6 tasks and second place on 1 dataset, which is the second-best model in general. CREPE is exceptionally strong at pitch- and note-related music tasks, such as NSynth 5h, NSynth 50h, and MAESTRO 5h, which most current speech/audio representation models fail to cope with. However, the authors note that CREPE's poor performances on other datasets can also be found. Wav2vec 2.0 and HuBERT, though only pretrained on speech corpus, are effective on various speech and non-speech tasks. In particular, for non-speech tasks, except our proposed ensemble framework, HuBERT achieves the second-best results on Gunshot Triangulation, Mridingham Tonic, and NSynth 5h. Nevertheless, their performances on the pitch- and note-related music tasks are not satisfying. Both wav2vec 2.0 and HuBERT fail to deal with MAESTRO 5h, and their performances on NSynth also substantially fall behind CREPE. Especially, HuBERT only gets 18.4 accuracy on NSynth 5h and 42.9 on NSynth 50h, which are the third from last among all methods. On the contrary, wav2vec 2.0 has over 20\% higher results than HuBERT on this task. This shows that representation models' feature extraction abilities on non-speech data are also affected by the design of their pretraining schema beside the training corpus.

It is also noteworthy that, though AudioSet and VGG-Sound are two large-scale datasets containing both speech and non-speech audio clips, models pre-trained on them, such as PaSST, PANNs, and Wav2CLIP, do not remarkably surpass other models that are only pre-trained on speech audio, such as SERAB BYOL-S, HuBERT xLarge and wav2vec 2.0 Large, or even images, such as EfficientNet-B2. In addition, models pre-trained on the same corpus can still have great performance gaps on testing datasets. For instance, PaSST base2levelmel acquires 54.1 Pitch accuracy (Acc) on NSynth 50h, while PANNs only get 30.1; wav2vec 2.0 Large has 65.3 Pitch accuracy on NSynth 50h, yet HuBERT xLarge only gets 42.9. This leads to a conclusion that, besides pre-training corpus, models' attention to input audio can also be influenced by other factors, such as model architectures and pre-training objectives. Even though existing SSL speech models like wav2vec 2.0 and HuBERT normally have similar Transformer-based structures, their abilities to extract speech and non-speech audio can still vary. This observation turns out to be the basis of the use of ensemble learning, and the promising performances of our proposed ensemble framework in turn validate this observation.

\begin{table}[hbtp]

  \caption{Performance of HuBERT xLarge and wav2vec 2.0 Large with and without feature fusion. Direct feature fusion over all hidden states increases SSL speech models' performances on most tasks, including speech and non-speech ones.}\label{Tab:3}
  \resizebox{\textwidth}{!} {
    
    \begin{tabular}{  l | c | c | c | c | c | c | c | c | c }
      
      \toprule
      
      \multicolumn{1}{r|}{\rotatebox{90}{\textbf{Task and measure}}} & \rotatebox{90}{\shortstack[l]{Bejing Opera\\ (Acc. $\uparrow$)}} & \rotatebox{90}{\shortstack[l]{CREMA-D\\(Acc. $\uparrow$)}} & \rotatebox{90}{\shortstack[l]{DCASE 2016\\(Onset FMS $\uparrow$)}} & \rotatebox{90}{\shortstack[l]{ESC-50\\(Acc. $\uparrow$)}} & \rotatebox{90}{\shortstack[l]{FSD50k\\(mAP $\uparrow$)}} & \rotatebox{90}{\shortstack[l]{GTZAN Genre\\(Acc. $\uparrow$)}} & \rotatebox{90}{\shortstack[l]{GTZAN Music/Speech\\(Acc. $\uparrow$)}} & \rotatebox{90}{\shortstack[l]{Gunshot\\(Acc. $\uparrow$)}} & \rotatebox{90}{\shortstack[l]{Libricount\\(Acc. $\uparrow$)}} \\
    
      \midrule
      
      wav2vec 2.0 Large  & {90.7} & {65.6} & {66.3} & {56.1} & {34.2} & {78.0} & {94.6} & {84.8} & \textBF{69.2} \\
      Fusion wav2vec 2.0 Large & \textBF{94.5} & \textBF{69.2} & \textBF{79.8} & \textBF{69.5} & \textBF{40.3} & \textBF{79.3} & \textBF{95.3} & \textBF{96.7} & {65.3} \\
      
      \midrule
      
      HuBERT xLarge & {94.5} & {69.0} & {58.4} & {60.3} & {31.4} & {73.5} & {91.3} & \textBF{93.2} & {64.6} \\
      Fusion HuBERT xLarge & \textBF{94.9} & \textBF{75.2} & \textBF{82.6} & \textBF{74.3} & \textBF{41.3} & \textBF{79.6} & \textBF{93.6} & {92.9} & \textBF{68.3} \\
      
      \bottomrule
      
      \toprule
      
      \multicolumn{1}{r|}{\rotatebox{90}{\textbf{Task and measure}}} & \rotatebox{90}{\shortstack[l]{Maestro 5h\\
    (Onset FMS $\uparrow$)}} & \rotatebox{90}{\shortstack[l]{Mridingham Stroke\\
    (Acc. $\uparrow$)}} & \rotatebox{90}{\shortstack[l]{Mridingham Tonic\\
    (Acc. $\uparrow$)}} & \rotatebox{90}{\shortstack[l]{NSynth Pitch 50h\\
    (Pitch Acc. $\uparrow$)}} & \rotatebox{90}{\shortstack[l]{NSynth Pitch 5h\\
    (Pitch Acc. $\uparrow$)}} & \rotatebox{90}{\shortstack[l]{Speech commands 5h\\
    (Acc. $\uparrow$)}} & \rotatebox{90}{\shortstack[l]{Speech commands full\\
    (Acc. $\uparrow$)}} & \rotatebox{90}{\shortstack[l]{Vocal Imitation\\
    (mAP $\uparrow$)}} & \rotatebox{90}{\shortstack[l]{VoxLingua107 top 10\\
    (Acc. $\uparrow$)}} \\

      \midrule
      
      wav2vec 2.0 Large & {03.3} & {94.3} & {82.8} & \textBF{65.3} & \textBF{40.2} & {83.8} & {87.9} & {08.0} & {49.3}  \\
      Fusion wav2vec 2.0 Large  & \textBF{11.1} & \textBF{96.2} & \textBF{83.8} & {60.6} & {33.0} & \textBF{95.7} & \textBF{96.9} & \textBF{17.4} & \textBF{70.6} \\
      
      \midrule
      
      HuBERT xLarge & {00.7} & {95.3} & {85.0} & {42.9} & {18.4} & \textBF{95.3} & {95.4} & {15.4} & {63.7} \\
      Fusion HuBERT xLarge & \textBF{16.6} & \textBF{97.4} & \textBF{90.9} & \textBF{68.8} & \textBF{38.2} & {94.7} & \textBF{95.7} & \textBF{18.5} & \textBF{71.4} \\
      
      \bottomrule
    \end{tabular}
  }
\end{table}

\begin{table}[hbtp]
\caption{Performance of feature averaging versus feature concatenation of wav2vec 2.0 Large plus HuBERT Large. Cat xw and avg xw denote the concatenation and the average of HuBERT Large and wav2vec 2.0 Large's representations, respectively.} \label{Tab:4}
\resizebox{\textwidth}{!}{

  \begin{tabular}{  l | c | c | c | c | c | c | c | c | c  }

    \toprule

    \multicolumn{1}{r|}{\rotatebox{90}{\textbf{Task and measure}}} & \rotatebox{90}{\shortstack[l]{Bejing Opera\\ (Acc. $\uparrow$)}} & \rotatebox{90}{\shortstack[l]{CREMA-D\\(Acc. $\uparrow$)}} & \rotatebox{90}{\shortstack[l]{DCASE 2016\\(Onset FMS $\uparrow$)}} & \rotatebox{90}{\shortstack[l]{ESC-50\\(Acc. $\uparrow$)}} & \rotatebox{90}{\shortstack[l]{FSD50k\\(mAP $\uparrow$)}} & \rotatebox{90}{\shortstack[l]{GTZAN Genre\\(Acc. $\uparrow$)}} & \rotatebox{90}{\shortstack[l]{GTZAN Music/Speech\\(Acc. $\uparrow$)}} & \rotatebox{90}{\shortstack[l]{Gunshot\\(Acc. $\uparrow$)}} & \rotatebox{90}{\shortstack[l]{Libricount\\(Acc. $\uparrow$)}} \\
    
    \midrule

    avg xw & \textBF{94.1} & \textBF{69.9} & {41.7} & \textBF{58.7} & {31.8} & \textBF{74.6} & {92.8} & {81.0} & {62.3} \\
    cat xw & {93.6} & {69.8} & \textBF{45.2} & {58.6} & \textBF{32.3} & {73.4} & \textBF{93.6} & \textBF{86.9} & \textBF{62.7} \\
        
    \bottomrule

    \toprule

    \multicolumn{1}{r|}{\rotatebox{90}{\textbf{Task and measure}}} & \rotatebox{90}{\shortstack[l]{Maestro 5h\\
    (Onset FMS $\uparrow$)}} & \rotatebox{90}{\shortstack[l]{Mridingham Stroke\\
    (Acc. $\uparrow$)}} & \rotatebox{90}{\shortstack[l]{Mridingham Tonic\\
    (Acc. $\uparrow$)}} & \rotatebox{90}{\shortstack[l]{NSynth Pitch 50h\\
    (Pitch Acc. $\uparrow$)}} & \rotatebox{90}{\shortstack[l]{NSynth Pitch 5h\\
    (Pitch Acc. $\uparrow$)}} & \rotatebox{90}{\shortstack[l]{Speech commands 5h\\
    (Acc. $\uparrow$)}} & \rotatebox{90}{\shortstack[l]{Speech commands full\\
    (Acc. $\uparrow$)}} & \rotatebox{90}{\shortstack[l]{Vocal Imitation\\
    (mAP $\uparrow$)}} & \rotatebox{90}{\shortstack[l]{VoxLingua107 top 10\\
    (Acc. $\uparrow$)}} \\
    
    \midrule
    
    avg xw & \textBF{00.4} & {94.6} & {79.9} & {41.5} & \textBF{19.8} & {90.7} & \textBF{95.2} & {16.5} & {69.0} \\
    cat xw & {00.3} & \textBF{95.1} & \textBF{80.2} & \textBF{42.8} & \textBF{19.8} & \textBF{93.6} & {92.9} & \textBF{16.6} & \textBF{70.6} \\
    
    \bottomrule
      
  \end{tabular}
}
\end{table}


        
        

\begin{table}[hbtp]

\caption{Performance of HuBERT xLarge with and without CREPE. Avg xc means HuBERT Large plus CREPE with simple feature averaging. Notice that incorporating CREPE can largely improve performances on music-related tasks, such as NSynth and MAESTRO. This demonstrates HuBERT's failure in music tasks as well as the effectiveness of the representation ensemble.}\label{Tab:5}
\resizebox{\textwidth}{!}{

  \begin{tabular}{  l | c | c | c | c | c | c | c | c | c  }

    \toprule

    \multicolumn{1}{r|}{\rotatebox{90}{\textbf{Task and measure}}} & \rotatebox{90}{\shortstack[l]{Bejing Opera\\ (Acc. $\uparrow$)}} & \rotatebox{90}{\shortstack[l]{CREMA-D\\(Acc. $\uparrow$)}} & \rotatebox{90}{\shortstack[l]{DCASE 2016\\(Onset FMS $\uparrow$)}} & \rotatebox{90}{\shortstack[l]{ESC-50\\(Acc. $\uparrow$)}} & \rotatebox{90}{\shortstack[l]{FSD50k\\(mAP $\uparrow$)}} & \rotatebox{90}{\shortstack[l]{GTZAN Genre\\(Acc. $\uparrow$)}} & \rotatebox{90}{\shortstack[l]{GTZAN Music/Speech\\(Acc. $\uparrow$)}} & \rotatebox{90}{\shortstack[l]{Gunshot\\(Acc. $\uparrow$)}} & \rotatebox{90}{\shortstack[l]{Libricount\\(Acc. $\uparrow$)}} \\
    
    \midrule

    HuBERT xLarge & \textBF{94.5} & \textBF{69.0} & {58.4} & \textBF{60.3} & \textBF{31.4} & \textBF{73.5} & {91.3} & \textBF{93.2} & \textBF{64.6} \\
    avg xc & {93.2} & {54.0} & \textBF{61.0} & {43.7} & {25.3} & {69.8} & \textBF{94.6} & {84.5} & {62.2} \\
        
    \bottomrule

    \toprule

    \multicolumn{1}{r|}{\rotatebox{90}{\textbf{Task and measure}}} & \rotatebox{90}{\shortstack[l]{Maestro 5h\\
    (Onset FMS $\uparrow$)}} & \rotatebox{90}{\shortstack[l]{Mridingham Stroke\\
    (Acc. $\uparrow$)}} & \rotatebox{90}{\shortstack[l]{Mridingham Tonic\\
    (Acc. $\uparrow$)}} & \rotatebox{90}{\shortstack[l]{NSynth Pitch 50h\\
    (Pitch Acc. $\uparrow$)}} & \rotatebox{90}{\shortstack[l]{NSynth Pitch 5h\\
    (Pitch Acc. $\uparrow$)}} & \rotatebox{90}{\shortstack[l]{Speech commands 5h\\
    (Acc. $\uparrow$)}} & \rotatebox{90}{\shortstack[l]{Speech commands full\\
    (Acc. $\uparrow$)}} & \rotatebox{90}{\shortstack[l]{Vocal Imitation\\
    (mAP $\uparrow$)}} & \rotatebox{90}{\shortstack[l]{VoxLingua107 top 10\\
    (Acc. $\uparrow$)}} \\
    
    \midrule
    
    HuBERT xLarge & {00.7} & \textBF{95.3} & \textBF{85.0} & {42.9} & {18.4} & \textBF{95.3} & \textBF{95.4} & \textBF{15.4} & \textBF{63.7}  \\
    avg xc & \textBF{46.2} & {91.7} & {83.1} & \textBF{89.7} & \textBF{87.8} & {73.7} & {82.3} & {07.9} & {29.3} \\
    
    \bottomrule
      
  \end{tabular}
}
\end{table}

\begin{table}[hbtp]

    \caption{Performance of wav2vec 2.0 Large plus CREPE with feature concatenation with and without HuBERT xLarge. Cat wc denotes the ensemble of wav2vec 2.0 Large plus CREPE with feature concatenation, while cat xwc means the same schema but with the join of HuBERT xLarge. Note that cat xwc generally outperforms cat wc, which means HuBERT xLarge provides extra acoustic information that makes the concatenated feature more holistic.}\label{Tab:6}
    \resizebox{\textwidth}{!}{
    
      \begin{tabular}{  l | c | c | c | c | c | c | c | c | c  }
    
        \toprule
    
        \multicolumn{1}{r|}{\rotatebox{90}{\textbf{Task and measure}}} & \rotatebox{90}{\shortstack[l]{Bejing Opera\\ (Acc. $\uparrow$)}} & \rotatebox{90}{\shortstack[l]{CREMA-D\\(Acc. $\uparrow$)}} & \rotatebox{90}{\shortstack[l]{DCASE 2016\\(Onset FMS $\uparrow$)}} & \rotatebox{90}{\shortstack[l]{ESC-50\\(Acc. $\uparrow$)}} & \rotatebox{90}{\shortstack[l]{FSD50k\\(mAP $\uparrow$)}} & \rotatebox{90}{\shortstack[l]{GTZAN Genre\\(Acc. $\uparrow$)}} & \rotatebox{90}{\shortstack[l]{GTZAN Music/Speech\\(Acc. $\uparrow$)}} & \rotatebox{90}{\shortstack[l]{Gunshot\\(Acc. $\uparrow$)}} & \rotatebox{90}{\shortstack[l]{Libricount\\(Acc. $\uparrow$)}} \\
        
        \midrule
    
        cat wc & {92.0} & {46.0} & {58.5} & {34.3} & {23.4} & {68.1} & \textBF{93.8} & {83.3} & {56.9} \\
        cat xwc & \textBF{96.6} & \textBF{74.2} & \textBF{82.6} & \textBF{73.4} & \textBF{42.0} & \textBF{80.5} & {92.8} & \textBF{93.5} & \textBF{69.7} \\
    
        \bottomrule
    
        \toprule
    
        \multicolumn{1}{r|}{\rotatebox{90}{\textbf{Task and measure}}} & \rotatebox{90}{\shortstack[l]{Maestro 5h\\
    (Onset FMS $\uparrow$)}} & \rotatebox{90}{\shortstack[l]{Mridingham Stroke\\
    (Acc. $\uparrow$)}} & \rotatebox{90}{\shortstack[l]{Mridingham Tonic\\
    (Acc. $\uparrow$)}} & \rotatebox{90}{\shortstack[l]{NSynth Pitch 50h\\
    (Pitch Acc. $\uparrow$)}} & \rotatebox{90}{\shortstack[l]{NSynth Pitch 5h\\
    (Pitch Acc. $\uparrow$)}} & \rotatebox{90}{\shortstack[l]{Speech commands 5h\\
    (Acc. $\uparrow$)}} & \rotatebox{90}{\shortstack[l]{Speech commands full\\
    (Acc. $\uparrow$)}} & \rotatebox{90}{\shortstack[l]{Vocal Imitation\\
    (mAP $\uparrow$)}} & \rotatebox{90}{\shortstack[l]{VoxLingua107 top 10\\
    (Acc. $\uparrow$)}} \\
        
        \midrule
    
        cat wc & \textBF{46.3} & {89.8} & {82.3} & \textBF{89.9} & \textBF{86.8} & {88.5} & {91.9} & {07.6} & {31.0}  \\
        cat xwc & {44.1} & \textBF{97.2} & \textBF{92.3} & {88.5} & {84.6} & \textBF{96.1} & \textBF{96.8} & \textBF{19.7} & \textBF{72.0}\\
    
        \bottomrule
          
      \end{tabular}
    }
\end{table}

\subsection{Ablation Study}
Ablation studies are conducted to verify the contributions of operations and components in our ensemble framework. Results are shown in \tableref{Tab:3,Tab:4,Tab:5,Tab:6}.

We first look into the use of intra-model feature aggregation. As displayed in \tableref{Tab:3}, the direct fusion of all hidden states helps models generate better representations. For instance, for DCASE 2016 Task 2, the event onset f-measure (FMS) of wav2vec 2.0 Large increases from 66.3 to 79.8 with the fusion strategy. HuBERT xLarge follows the same trend but with more significant improvements, with event onset FMS rocketing to 82.6. However, feature aggregation does not always produce better results. For example, wav2vec 2.0 Large's classification accuracy on LibriCount drops from 69.2 to 65.3 after applying the feature fusion. Nevertheless, this technique still generally has a positive effect on models' representation ability.
 As for feature concatenation versus feature averaging, \tableref{Tab:3} shows that the cat xw has better performances on eleven tasks and avg xw wins seven, with one tie. With the abundant results shown on the 18 tasks, we conclude that, by and large, feature concatenation brings more benefits than feature averaging, and so we adopt the former to our framework. \tableref{Tab:5} illustrates the effects of adding CREPE. With CREPE, HuBERT xLarge's performances on DCASE 2016 Task 2, GTZAN Music/Speech, MAESTRO 5h, NSynth 50h, and NSynth 5h rise, especially significantly in NSynth 50h, NSynth 5h, and MAESTRO 5h. The four tasks are all music-related. Besides, DCASE 2016 Task 2 and MAESTRO 5h are event onset detection datasets. It can be derived that, SSL speech models may still have blind spots in music-related content, and their abilities on event detection datasets are also not as strong as in other kinds of tasks such as classification/identification. Though incorporating CREPE solves the issue, it leads to drops in other tasks by a considerable margin. For instance, its performance on Speech Commands full decreases from 95.4 accuracy to 82.3 after combining CREPE. The reason is that, given that embeddings on Speech Command full produced by CREPE contain very little content information, it will further jeopardize the information in embeddings generated by HuBERT Large if we take the average of these two vectors. The solution is to use feature concatenation rather than averaging, and our concatenation-based final framework indeed achieves 96.8 accuracy on Speech Command full even with CREPE being inside. Finally, the contribution of HuBERT xLarge among all three models is examined in \tableref{Tab:6}. It is shown that HuBERT xLarge helps cat xwc attain higher results on nearly all tasks. For example, the event onset FMS rises from 58.5 to 82.6 in DCASE 2016 Task 2, and accuracy from 34.3 to 73.4 in ESC-50.

\section{Conclusion}
This work proposes a representation ensemble framework that integrates models' audio representations and leverages the framework to investigate SSL speech models' effectiveness on speech and non-speech datasets. Experimental results show that SSL speech models' are capable of extracting meaningful representations of various non-speech corpus, such as instrument classification, while failure on fine-grained music-based tasks, such as pitch- and note-related datasets is also observed. In addition, different SSL speech models might have insights into different aspects of audio features. Specifically, wav2vec 2.0 and HuBERT are shown to have considerable performance gaps on some datasets, such as NSynth Pitch and VoxLingua107. We also investigate the contribution of each component, such as feature concatenation, of the ensemble framework, and the overall performance of our method generally surpasses its state-of-the-art constituent models and other teams' methods in HEAR Challenge.











\bibliography{pmlr176/main,pmlr176/pmlr176}

\end{document}